\documentclass[12pt]{article}
\usepackage{graphics}
\usepackage{epsfig}
\newcommand{\be}{\begin{eqnarray}}
\newcommand{\en}{\end{eqnarray}}

\newcommand{\nn}{\nonumber}

\begin{document}

\begin{center}
{\Large \bf One-Loop Helicity Amplitudes for Parton Level \\
Virtual Compton Scattering}
 
\vskip 1\baselineskip

H.W.\ Huang$^a$
%\footnote{Postdoctoral research fellow (No. P99221) of
%the Japan Society for the Promotion of Science (JSPS).},
and T. Morii$^b$\\[0.5em]
{\small{\it $a$ Department of Physics, University of Colorado, Boulder, 
CO 80309-0390, USA}}\\
{\small{\it $b$ Faculty of Human Development, Kobe University, Nada,
Kobe 657-8501, Japan}}
\vskip \baselineskip
\end{center}

\vskip 3\baselineskip

\begin{abstract}
We calculate the one-loop QCD virtual corrections to all helicity
amplitudes for parton level virtual Compton scattering processes. We
include the amplitudes both on quark target process $\gamma^*
q\rightarrow\gamma q$ and on gluon target process 
$\gamma^*g\rightarrow\gamma g$. The infrared pole structure of the
amplitudes is in agreement with the prediction of Catani's general
formalism for the singularities of one-loop amplitudes, while
expressions for the finite remainder are given in terms of logarithms
and dilogarithms that are real in the physical region.
\end{abstract}

\newpage

\section{Introduction}
Photons, being real or virtual, are known to be clean probes of the internal
structure of the nucleon. One of the important processes of photon is
Compton scattering, which refers to elastic scattering of a photon off
a charged object. The Compton reaction in different
kinematical situations can provide different physical information
about the behavior of quarks and gluons in the nucleon. There are 
two complementary kinematical regions, which are deeply virtual 
region and wide-angle
region. The region of deeply virtual Compton scattering (DVCS) is
characterized by small momentum transfer from the initial to final
nucleon and a large photon virtuality, while in the region of
wide-angle Compton scattering (WACS) the situation is reversed.

The theoretical development on DVCS has revealed that in the Bjorken
limit, this process is dominated by the simple handbag mechanism in
which a quark (an antiquark) in the initial nucleon absorbs the
virtual photon, immediately radiates a real photon, and falls back to
form the recoiled nucleon \cite{Dittes:xz,Muller:1998fv,ji97}. It has
been proved that DVCS amplitude can be factorized into the finite
perturbative parts and the nonperturbative parts represented by
collinearly-divergent terms, which correspond to the matrix elements
of a class of newly introduced generalized parton distributions (GPDs)
\cite{Radyushkin:1997ki,Ji:1998xh,Collins:1998be}. These GPDs are
hybrid objects, which combine properties of form factors and of
ordinary parton distributions.
 
For WACS process, there is a general agreement that in
asymptotic large momentum transfer region, the amplitude for Compton
scattering is dominated by Brodsky--Lepage hard scattering picture
\cite{brodsky}, which is given by the convolution of a hard-scattering
amplitude of collinear constituent partons and the distribution amplitude
of hadrons. The perturbative contribution has been calculated in
\cite{farrar,kron91,van97} to leading twist accuracy. However, the
cross sections predicted in that approach are way below the existing
Compton data, unless strongly asymmetric, i.e. end-point concentrated
distribution amplitudes are used. This results in the conclusion that 
the cross sections are
dominated by contributions from the soft end-point regions, where the
assumptions of the leading twist perturbative calculation break
down. Thus, the hard scattering model of Brodsky--Lepage, although
likely to be the true asymptotic picture for exclusive reactions,
does not seem to be dominant at moderately large momentum transfer, the
kinematical region being accessible to present-day
experiments. 

Recently, a new mechanism has been introduced, in which the physics of
handbag diagram is also of importance even for WACS reaction. As has
been argued in \cite{Radyushkin:1998rt,diehl,Diehl:1999tr}, at
moderately large momentum transfer, Compton scattering off protons
approximately factorizes into a hard parton--photon subprocess and a
soft-proton matrix element described by new form factors specific to
Compton scattering. These new form factors represent moments of GPDs
and can be modelled by overlaps of light-cone wave functions, which
provide the link between exclusive and inclusive reactions. Using
light-cone wave function overlaps as a model of the GPDs, detailed
predictions of cross sections and polarization observables for real
and virtual Compton scattering have been achieved in
\cite{diehl,Diehl:1999tr} by using the leading order (LO) results for
the hard parton--photon subprocess.  In addition, the perturbative
contributions appeared in the factorization formalisms for both DVCS
and WACS processes have been calculated to one-loop level so far. The
authors of
\cite{Ji:1998xh,Belitsky:1997rh,Mankiewicz:1997bk,Belitsky:1998wz}
have considered the one-loop corrections to DVCS process, while the
next-to-leading order (NLO) QCD corrections to the parton level
subprocess of real WACS processes have been calculated in
\cite{morii}. Some numerical predictions were also presented in the
latter paper \cite{morii} based on the GPDs model proposed in
\cite{diehl}.

The purpose of the present paper is to
calculate the NLO corrections to the parton level virtual Compton
scattering in wide-angle region. Our results together with the GPDs
form factors will provide the necessary ingredients for calculating
the wide-angle virtual Compton scattering at the NLO. These results
may also be used to compare with other theoretical results in order to
test the handbag mechanism and facilitate the interpretation of future
experimental data that might be obtained at Jefferson Lab or at an
ELFE-type accelerator at DESY or CERN.

The paper is organized as follows: Starting with the kinematical
discussion, we calculate the LO amplitudes in Sect.2. Detailed
calculations about the NLO results are given in Sect.3. Sect. 4 is
devoted to the check of our calculations by comparing the divergent
parts of our formulas with Catani's general formulas, and also by
comparing the expressions of our formulas in the $-t\rightarrow 0$
(DVCS) and $Q^2\rightarrow 0$ (real WACS) limits with those results
appeared in
\cite{Ji:1998xh,Belitsky:1997rh,Mankiewicz:1997bk,Belitsky:1998wz,morii,dixon}. The
paper terminates with a few concluding remarks (Sect. 5).

\section{LO amplitudes}
For the parton level virtual Compton process
\begin{equation}
\gamma^*(q)+parton(p)\longrightarrow\gamma(q^\prime)+parton(p^\prime),
\end{equation}
we work in the center of mass system (CMS) of photon and parton. Neglecting
the parton mass the different particle momenta can be chosen to be
\begin{eqnarray}\nonumber
q^{\mu}&=&(q_0,0,0,|{\bf{p}}|),\\\nonumber
p^{\mu}&=&|{\bf{p}}|(1,0,0,-1),\\\nonumber
q^{\prime\mu}&=&|{\bf{p}^{\prime}}|(1,\sin\theta,0,\cos\theta),
\\\label{kinematic}
p^{\prime\mu}&=&|{\bf{p}^{\prime}}|(1,-\sin\theta,0,-\cos\theta),
\end{eqnarray}
in that frame. Here $\theta$ is the scattering angle of the outgoing real
photon. $q_0,|\bf{p}|,|\bf{p}^\prime|$ and $\theta$ are related to the
Mandelstam variables $s=(p+q)^2,t=(p^\prime-p)^2$ and to the
virtuality $(q^2=-Q^2)$ of the incoming photon by
\begin{eqnarray}\nonumber
q_0=\frac{s-Q^2}{2\sqrt{s}},&&|{\bf{p}}|=\frac{s+Q^2}{2\sqrt{s}},
\\
|{\bf{p}}^{\prime}|=\frac{\sqrt{s}}{2},&&\cos\theta=1+\frac{2t}{s+Q^2}.
\end{eqnarray}
In CMS, the polarization vectors of incoming virtual photon are
\begin{equation}
\epsilon_q(\pm)=\mp\frac{1}{\sqrt{2}}(0,1,\pm i,0),~~
\epsilon_q(0)=\frac{1}{Q}(|{\bf{p}}|,0,0,q_0).
\end{equation}
The polarization vector of the outgoing real photon
is given by
\begin{equation}
\epsilon_{q\prime}(\mu^\prime)=\frac{1}{\sqrt{2}}(0,-\mu^\prime \cos\theta,-i,
\mu^\prime \sin\theta).
\end{equation}

\begin{figure} 
\begin{center}
\epsfig{figure=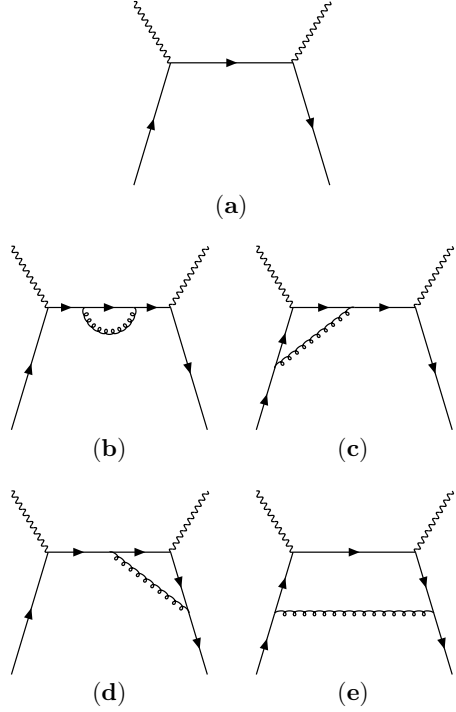, bb= 215 405 430 690, width=7.0cm}
\end{center}
\vspace{-0.4cm}
\caption{\label{fig1}Feynman graphs for Compton scattering off
on-shell quarks. a) is the LO graph, the others represent the NLO QCD
corrections. Graphs with self-energy corrections to external fermions
and those with interchanged interaction points of the photons are not
shown.}
\end{figure}

The tree amplitudes only receive contribution from scattering with
quark. There are two Feynman diagrams which contribute to the LO
amplitude for the reaction $\gamma^*q\rightarrow\gamma q$; a half of
them is shown in Fig.\ref{fig1}(a). The other half will be taken into
account by using the crossing symmetry, i.e. the simultaneous
replacement of $q\leftrightarrow -q^\prime$ and
$\epsilon_q\leftrightarrow\epsilon^*_{q^\prime}$. The LO amplitude is
\begin{eqnarray}
{\cal M}^{(0)}_{\mu^\prime\lambda^\prime,\mu\lambda}&=&
e_q^2e^2M^{(0)}_{\mu^\prime\lambda^\prime,\mu\lambda},
\en
where $e_q$ is the electric charge of quark $q$, and $\mu(\lambda)$ and
$\mu^\prime(\lambda^\prime)$ denote the helicity of the
initial and final state photons (partons), respectively. The amplitude 
$M^{(0)}_{\mu^\prime\lambda^\prime,\mu\lambda}$ is written as
\be
M^{(0)}_{\mu^\prime\lambda^\prime,\mu\lambda}&=&
\bar{u}(p^\prime,\lambda^\prime)\left[\rlap/{\epsilon}^*_{q^\prime}(\mu^\prime)
\frac{\rlap/{p}+\rlap/{q}}{(p+q)^2+i\epsilon}\rlap/{\epsilon}_q(\mu)
+\rlap/{\epsilon}_q(\mu)\frac{\rlap/{p}-\rlap/{q}^\prime}
{(p-q^\prime)^2+i\epsilon}\rlap/{\epsilon}_{q^\prime}^*(\mu^\prime)\right]
u(p,\lambda).
\end{eqnarray}
Working out in the photon--parton CMS, we explicitly find
\begin{eqnarray}\label{LO}\nn
M^{(0)}_{++,++}=2\sqrt{\frac{s}{-u}}\frac{s+Q^2}{s},&&
M^{(0)}_{-+,-+}=2\sqrt{\frac{-u}{s}}\frac{s}{s+Q^2},\\\nn
M^{(0)}_{-+,++}=2\frac{Q^2}{s+Q^2}\frac{t}{\sqrt{-su}},&&
M^{(0)}_{++,-+}=0,\\ M^{(0)}_{-+,0+}=2\frac{Q}{s+Q^2}\sqrt{-2t},&&
M^{(0)}_{++,0+}=0.  \en For the sake of legibility, explicit helicities
are labeled only by their signs. Since the quarks are taken as
massless, there is no quark helicity flip amplitude, i.e.
$M_{\mu^\prime-\lambda,\mu\lambda}=0$, to any order of
$\alpha_s$. Other helicity amplitudes can be obtained from those given in
(\ref{LO}) by parity invariance:
\begin{equation}
M_{-\mu^\prime-\lambda^\prime,-\mu-\lambda}
=(-1)^{\mu-\lambda-\mu^\prime+\lambda^\prime}
M_{\mu^\prime\lambda^\prime,\mu\lambda}.
\end{equation}

\section{One loop processes}

\begin{figure}[t]
\begin{center}
\epsfig{figure=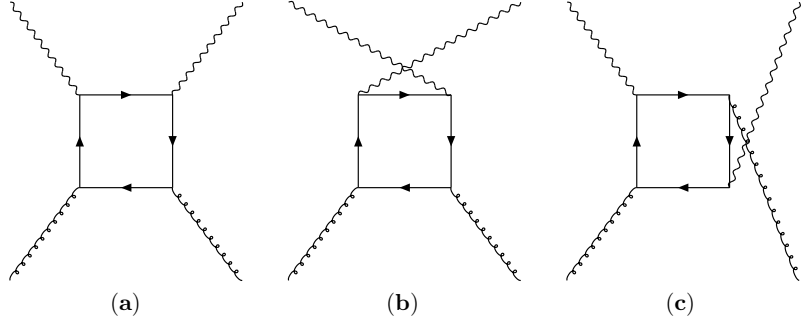, bb= 140 550 490 690, width=10.8cm}
\end{center}
\caption{\label{fig3} Sample Feynman graphs for photon--gluon scattering.} 
\end{figure}

The NLO corrections to $\gamma^*q\rightarrow\gamma q$ can be
calculated from the Feynman graphs (b)-(e) depicted in
Fig.\ref{fig1}. We work in Feynman gauge and use dimensional
regularization ($n=4-2\epsilon$). There are two mitigating factors
which simplify the NLO corrections. The first is that the calculation
does not contain ultraviolet (UV) singularities since the graphs 
in Fig. \ref{fig1} do not contribute to the renormalization of the 
strong, electromagnetic, or weak coupling constants. The second is that 
the self-energy insertions on the external quark lines vanish due to 
the cancellation of the UV and infrared (IR) divergences \cite{neerven}. 
Basically, what happens is that the UV and IR poles cancel when one does not
distinguish between them.

We evaluate these amplitudes using the Feynman parametrization
technique. The loop integrals associated with the four-point function
from the box diagrams shown in Fig.\ref{fig1}(e) are very difficult
to evaluate when powers of the loop momenta appear in the
numerator. However, one can express these tensor integrals in terms of
lower rank tensor integrals with the same number of propagators and
lower rank tensors with fewer propagators \cite{veltman}. In the end,
the four-point functions with powers of the loop momentum in the
numerator are reduced to a four-point functions with a constant
numerator and three- and two-point functions which are easier to
evaluate. The all loop integrals can be reduced to a set of 13 scalar
integrals which are given in Appendix.

The NLO corrections to the $\gamma^*q\rightarrow\gamma q$ amplitudes
read 
\be\nn
M^{q(1)}_{++,++}&=&M^{(0)}_{++,++}\frac{\alpha_s}{\pi}C_Ff(\epsilon)
\left\{-\frac{1}{(-t)^\epsilon}\left(\frac{1}{2\epsilon^2}
+\frac{3}{4\epsilon}\right)-\frac{9}{4}+\frac{\pi^2}{12}\right.\\\nn
&&+\frac{(2su-tQ^2)(s+u)+2(s+t)su}{4(s+t)(s+u)(t+u)^2}t
-\frac{(4st+su+4t^2+2tu)su}{4(s+t)^2(t+u)^2}{\rm ln}\frac{Q^2}{-u}
\\\label{quark_pp}
&&\left.-\left(\frac{3}{4}+\frac{(2Q^2+t)stu}{2(s+u)^2(t+u)^2}\right)
{\rm ln}\frac{-t}{Q^2}
-\frac{f_1(s,t,u)}{4}-\frac{t^2}{4(t+u)^2}f_2(s,t,u)\right\},\\\nn
M^{q(1)}_{-+,-+}&=&M^{(0)}_{-+,-+}\frac{\alpha_s}{\pi}C_Ff(\epsilon)
\left\{-\frac{1}{(-t)^\epsilon}\left(\frac{1}{2\epsilon^2}
+\frac{3}{4\epsilon}\right)-\frac{9}{4}+\frac{\pi^2}{12}
-\frac{t}{2(s+u)}\right.\\\nn
&&+\frac{2t-u}{4u}\left({\rm ln}\frac{-t}{s}+i\pi\right)+\frac{(tu+st-uQ^2)Q^2}
{2(s+u)^2u}{\rm ln}\frac{-t}{Q^2}\\\label{quark_mm}
&&\left.-\frac{t^2}{4u^2}f_1(s,t,u)-\frac{f_2(s,t,u)}{4}\right\},\\\nn
M^{q(1)}_{-+,++}&=&M^{(0)}_{-+,++}\frac{\alpha_s}{\pi}C_Ff(\epsilon)
\left\{-\frac{1}{(-t)^\epsilon}\left(\frac{1}{2\epsilon^2}
+\frac{3}{4\epsilon}\right)-\frac{9}{4}+\frac{\pi^2}{12}
-\frac{3s^2+4su+3u^2}{4(s+u)^2}{\rm ln}\frac{-t}{Q^2}\right.\\
&&\left.-\frac{f_1(s,t,u)+f_2(s,t,u)}{4}\right\}-\frac{\alpha_s}{2\pi}C_F
\frac{[(s+u)(t+u)+2su]t}{\sqrt{-su}(s+u)(s+Q^2)},\\\nn
M^{q(1)}_{++,-+}&=&\frac{\alpha_s}{\pi}C_F\sqrt{\frac{-u}{s}}\frac{t}{s+Q^2}
\left\{\left(\frac{2s+u}{(s+u)^2}{\rm ln}\frac{Q^2}{-t}+\frac{3s+2t}{2(s+t)^2}
{\rm ln}\frac{Q^2}{-u}-\frac{f_2(s,t,u)}{2s}\right)Q^2\right.\\
&&\left.+\frac{s-u-2Q^2}{2(s+t)(s+u)}s \right\},\\\nn
M^{q(1)}_{-+,0+}&=&M^{(0)}_{-+,0+}\frac{\alpha_s}{\pi}C_Ff(\epsilon)
\left\{-\frac{1}{(-t)^\epsilon}\left(\frac{1}{2\epsilon^2}
+\frac{3}{4\epsilon}\right)-\frac{9}{4}+\frac{\pi^2}{12}
-\frac{1}{2}\left({\rm ln}\frac{-t}{s}+i\pi\right)\right.\\
&&\left.+\frac{u}{2(s+u)}-\frac{s^2-2ut-u^2}{4(s+u)^2}{\rm ln}\frac{-t}{Q^2}
-\frac{f_2(s,t,u)}{4}+\frac{t}{4u}f_1(s,t,u)\right\},\\\nn
M^{q(1)}_{++,0+}&=&\frac{\alpha_s}{\pi}C_F\frac{Q\sqrt{-2t}}{s+Q^2}
\left\{\frac{st-u^2-su}{(s+u)^2}{\rm ln}\frac{-t}{Q^2}+\frac{2st-su+2t^2}{2(s+t)^2}
{\rm ln}\frac{-u}{Q^2}\right.\\\label{quark_0p}
&&\left.+\frac{t}{2s}f_2(s,t,u)+\frac{st-2su-tu}{2(s+t)(s+u)}\right\},
\en 
where the functions $f(\epsilon)$ and $f_i(s,t,u),i=1,2,3$ are all
defined in the Appendix. We see that the NLO amplitudes possess both
non-zero imaginary parts and non-zero photon helicity flips.

At the one-loop level, there is a complication which we have to
discuss next, namely gluons have to be considered as active partons as
well. In contrast to the case of quarks, the partonic amplitudes now
allow parton, i.e. gluon helicity flips to occur. The photon--gluon
amplitudes can be calculated from the three graphs shown in
Fig. \ref{fig3}. There are three further graphs contributing to order
$\alpha_s$ which however reduce to the first three ones by reversing
the fermion number flow. The fermion loop contributions satisfy the
color decomposition
\be
{\cal M}^{g(1)}_{a^\prime\mu^\prime\lambda^\prime,a\mu\lambda}&=&
\sum_qe_q^2e^2\delta_{aa^\prime}M^{g(1)}_{\mu^\prime\lambda^\prime,\mu\lambda},
\en
where $a_1$ and $a_2$ are color indices of incoming and outgoing gluons.
After some algebra, we find for the gluon helicity non-flip amplitudes
\be\nn
M^{g(1)}_{++,++}&=&
\frac{\alpha_s}{\pi}\left\{-\frac{t^2+u^2}{2s(t+u)}f_2(s,t,u)
+\frac{(st-2su-2u^2)t}{(s+u)^2(t+u)}{\rm ln}\frac{Q^2}{-t}\right.
\\\label{gluon_pp}
&&\left.-\frac{(2st-su+2t^2)u}{(s+t)^2(t+u)}{\rm ln}\frac{Q^2}{-u}
+\frac{(s-Q^2)tu}{(s+t)(s+u)(t+u)}\right\},\\\nn
M^{g(1)}_{-+,-+}&=&\frac{\alpha_s}{\pi}\left\{-\frac{(2t^2+2tu+u^2)s}{2u^2(t+u)}
f_1(s,t,u)+\frac{(2t+u)s}{(t+u)u}\left({\rm ln}\frac{-t}{s}+i\pi\right)\right.\\
&&\left.-\frac{(2st+su+3tu+u^2)sQ^2}{(s+u)^2(t+u)u}{\rm ln}\frac{Q^2}{-t}
-\frac{st}{(s+u)(t+u)}\right\},\\
M^{g(1)}_{-+,++}&=&\frac{\alpha_s}{\pi}\left\{\frac{tQ^2}{2u(t+u)}f_1(s,t,u)-
\frac{tuQ^2}{(s+u)^2(t+u)}{\rm ln}\frac{Q^2}{-t}+\frac{st}{(s+u)(t+u)}\right\},\\\nn
M^{g(1)}_{++,-+}&=&\frac{\alpha_s}{\pi}\left\{\frac{utQ^2}{s^2(t+u)}f_2(s,t,u)
-\frac{(3s+2u)tuQ^2}{(s+u)^2(t+u)s}{\rm ln}\frac{Q^2}{-t}\right.\\
&&\left.-\frac{(3s+2t)tuQ^2}{(s+t)^2(t+u)s}{\rm ln}\frac{Q^2}{-u}
-\frac{(s-Q^2)tu}{(s+t)(t+u)(s+u)}\right\},\\\nn
M^{g(1)}_{++,0+}&=&\frac{\alpha_s}{\pi}\left\{-\sqrt{\frac{ut}{2s}}\frac{(t-u)Q}
{s(t+u)}f_2(s,t,u)+\sqrt{\frac{2ut}{s}}\frac{(2st-su+tu-u^2)Q}{(s+u)^2(t+u)}
{\rm ln}\frac{Q^2}{-t}\right.\\
&&\left.+\sqrt{\frac{2ut}{s}}\frac{(st-2su+t^2-tu)Q}{(s+t)^2(t+u)}{\rm ln}\frac{Q^2}{-u}
-\frac{\sqrt{2uts}(t-u)Q}{(s+t)(s+u)(t+u)}\right\},\\\nn
M^{g(1)}_{-+,0+}&=&\frac{\alpha_s}{\pi}\left\{\sqrt{\frac{uts}{2}}\frac{tQ}
{u^2(t+u)}f_1(s,t,u)-\frac{\sqrt{2uts}Q}{(t+u)u}\left({\rm ln}\frac{-t}{s}+i\pi\right)
\right.\\
&&\left.+\frac{\sqrt{2uts}(s+2u)Q^3}{(s+u)^2(t+u)u}{\rm ln}\frac{Q^2}{-t}
+\frac{\sqrt{2uts}Q}{(s+u)(t+u)}\right\},  
\en 
and for the gluon helicity flip amplitudes
\be 
M^{g(1)}_{++,+-}&=&\frac{\alpha_s}{\pi},\\
M^{g(1)}_{-+,--}&=&\frac{\alpha_s}{\pi}\left(\frac{uQ^2}{2t(t+u)}f_3(s,t,u)+\frac{su}{(s+t)(t+u)}
-\frac{tuQ^2}{(s+t)^2(t+u)}{\rm ln}\frac{Q^2}{-u}\right),\\
M^{g(1)}_{-+,+-}&=&\frac{\alpha_s}{\pi}\left(-\frac{(t^2+2tu+2u^2)s}{2t^2(t+u)}f_3(s,t,u)-\frac{su}{(s+t)(t+u)}\right.
\\\nn &&\left.+\frac{s(t+2u)}{(t+u)t}\left({\rm ln}\frac{Q^2}{s}+i\pi\right)
+\frac{su(2st+2su+2t^2+3tu)}{(s+t)^2(t+u)t}{\rm ln}\frac{Q^2}{-u}\right),\\
M^{g(1)}_{++,--}&=&\frac{\alpha_s}{\pi},\\ 
M^{g(1)}_{++,0-}&=&0,\\\nn
M^{g(1)}_{-+,0-}&=&\frac{\alpha_s}{\pi}\left(\sqrt{\frac{uts}{2}}
\frac{uQ}{t^2(t+u)}f_3(s,t,u)+\frac{\sqrt{2uts}Q}{(s+t)(t+u)}\right.\\\label{gluon_0m}
&&\left.-\frac{\sqrt{2uts}Q}{(t+u)t}\left({\rm ln}\frac{Q^2}{s}+i\pi\right)
-\frac{\sqrt{2uts}Q(st+su+t^2+2tu)}{(s+t)^2(t+u)t}
{\rm ln}\frac{Q^2}{-u}\right).
\en 
We find that for all helicity non-flip amplitudes, the scaling
behavior of $-t\rightarrow 0$ are satisfied, which is demanded by the
angular momentum conservation.

\section{Comparison with other results}

We find that those helicity amplitudes which are non-zero at LO, turn
out to include infrared divergent parts which go as an universal form
\be
\label{factor} -M^{(0)}_{\mu^\prime\lambda^\prime,\mu\lambda}
\frac{\alpha_s}{2\pi}C_Ff(\epsilon)
\left(\frac{1}{\epsilon^2}+\frac{3}{2\epsilon}\right)\frac{1}{(-t)^\epsilon},
\en 
where the $1/\epsilon^2$ term appears as a consequence of overlapping
soft and collinear divergences. Catani presented a general
formula for the structure of infrared divergences of any QCD amplitude
\cite{catani}. For the case of the one-loop amplitude, Catani's
formula is 
\be\label{divergent}
M^{(0)}\frac{1}{2}\frac{-e^{\epsilon\psi(1)}}{\Gamma(1-\epsilon)}
\sum_i\frac{1}{{{\bf T}_i}^2}\left({{{\bf T}_i}}^2\frac{1}{\epsilon^2}
+\gamma_i\frac{1}{\epsilon}\right)\sum_{j\neq i}{{\bf T}_i}\cdot{{\bf
T}_j} \left(\frac{\mu^2e^{-i\lambda_{ij}\pi}}{2p_i\cdot
p_j}\right)^\epsilon \frac{\alpha_sS_\epsilon}{2\pi}, 
\en 
where $S_\epsilon$ is the typical phase-space volume factor in
$d=4-2\epsilon$ ($\gamma_E=-\psi(1)=0.5772\cdots$ is the Euler number)
\be 
S_\epsilon={\rm exp}[\epsilon({\rm ln}4\pi+\psi(1))].  
\en 
In (\ref{divergent}), the sum is over all external partons of the
amplitudes. $\lambda_{ij}=+1$ if $i$ and $j$ are both incoming or
outgoing partons and $\lambda_{ij}=0$ otherwise. The color charge
${\bf T}_i=\{T^a_i\}$ is a vector with respect to the color indices
$a$ of the emitted gluon, and an $SU(N)$ matrix with respect to the color
indices of the parton $i$. Applying (\ref{divergent}) to the process
we consider here, one may convert two of the partons to photons by
setting ${{\bf T}_i}\cdot{\bf T_\gamma}\rightarrow 0$ and ${{\bf
T}_i}\cdot{{\bf T}_j}=-{{\bf T}_i}^2$ to obtain the simplified formula
\be
-M^{(0)}\frac{\alpha_s}{2\pi}\frac{{\rm exp}(\epsilon {\rm
ln}4\pi)}{\Gamma(1-\epsilon)}
C_F\left(\frac{1}{\epsilon^2}+\frac{3}{2\epsilon}\right)
\left(\frac{\mu^2}{-t}\right)^\epsilon, 
\en 
the divergent part of which is exactly equal to the one given in 
eq.(\ref{factor}). This agreement provides a stringent check of 
our amplitudes.

To perform further checks on the reliability of amplitudes, 
we compare the amplitudes of real photon--parton
scattering processes obtained from our formulas with those already
existed results. The helicity amplitudes with one-loop corrections to
real Compton process $\gamma q\rightarrow\gamma q$ can be derived from
(\ref{quark_pp})--(\ref{quark_0p}) by taking $Q^2=0$, and they are
\be\nn
M^{q(1)}_{+,++}&=&f(\epsilon)
\frac{C_F\alpha_s}{\pi}\left\{-\frac{1}{(-t)^\epsilon}
\left(\frac{1}{\epsilon^2}+\frac{3}{2\epsilon}\right)-\frac{7}{2}
+\frac{\pi^2}{6}\right.\\
&+&\left.\frac{2t-s}{2s}{\rm ln}\frac{t}{u}+\frac{1}{2}{\rm ln}^2\frac{-t}{s}
+\frac{t^2}{2s^2}\left({\rm ln}^2\frac{t}{u}+\pi^2\right)+i\pi {\rm ln}\frac{-t}{s}\right\}
\sqrt{\frac{s}{-u}},\\\nonumber
M^{q(1)}_{+,--}&=&f(\epsilon)\frac{C_F\alpha_s}{\pi}
\left\{-\frac{1}{(-t)^\epsilon}\left(\frac{1}{\epsilon^2}
+\frac{3}{2\epsilon}\right)-\frac{7}{2}
+\frac{\pi^2}{6}\right.\\\nn
&+&\frac{2t-u}{2u}{\rm ln}\frac{-t}{s}+\frac{1}{2}\left(
{\rm ln}^2\frac{t}{u}+\pi^2\right)
+\frac{t^2}{2u^2}{\rm ln}^2\frac{-t}{s}\\
&+&\left.+i\pi\left(\frac{2t-u}{2u}
+\frac{t^2}{u^2}{\rm ln}\frac{-t}{s}\right)\right\}\sqrt{\frac{-u}{s}},\\
M^{q(1)}_{+,+-}&=&-\frac{C_F\alpha_s}{2\pi}\left(\sqrt{\frac{s}{-u}}
+\sqrt{\frac{-u}{s}}\right),\\
M^{(1)}_{+,-+}&=&-\frac{C_F\alpha_s}{2\pi}\left(\sqrt{\frac{s}{-u}}
+\sqrt{\frac{-u}{s}}\right),\\
M^{q(1)}_{+,0+}&=&0\\
M^{q(1)}_{+,0-}&=&0.
\end{eqnarray}

We can see that the contributions from longitudinal photon disappear and
all other amplitudes agree with those given in \cite{morii}. Applying
$Q^2=0$ to the real photon--gluon scattering process $\gamma
g\rightarrow\gamma g$, (\ref{gluon_pp})--(\ref{gluon_0m}) can be
simplified as
\be
M^{g(1)}_{++,++}&=&-\frac{\alpha_s}{\pi}\left\{\frac{t^2+u^2}{2s^2}
\left({\rm ln}^2\frac{t}{u}+\pi^2\right)
+\frac{s+2t}{s}{\rm ln}\frac{t}{u}+1\right\},\\
M^{g(1)}_{++,--}&=&-\frac{\alpha_s}{\pi}\left\{\frac{s^2+t^2}{2u^2}{\rm ln}^2
\frac{-t}{s}+\frac{t-s}{u}{\rm ln}
\frac{-t}{s}+1+i\pi\left(\frac{t-s}{u}+\frac{s^2+t^2}{u^2}
{\rm ln}\frac{-t}{s}\right)\right\},\\
M^{g(1)}_{++,+-}&=&\frac{\alpha_s}{\pi},\\
M^{g(1)}_{++,-+}&=&\frac{\alpha_s}{\pi},\\
M^{g(1)}_{++,0+}&=&0,\\
M^{g(1)}_{++,0-}&=&0,\\
M^{g(1)}_{+-,++}&=&\frac{\alpha_s}{\pi},\\
M^{g(1)}_{+-,--}&=&\frac{\alpha_s}{\pi},\\\nn
M^{g(1)}_{+-,+-}&=&-\frac{\alpha_s}{\pi}\left\{\frac{s^2+u^2}{2t^2}{\rm ln}^2
\frac{-u}{s}+\frac{u-s}{t}{\rm ln}\frac{-u}{s}+1\right.\\
&&\left.+i\pi\left(\frac{u-s}{t}
+\frac{s^2+u^2}{t^2}{\rm ln}\frac{-u}{s}\right)\right\},\\
M^{g(1)}_{+-,-+}&=&\frac{\alpha_s}{\pi},\\
M^{g(1)}_{+-,0+}&=&0,\\
M^{g(1)}_{+-,0-}&=&0,
\en
which are also equal to the results obtained in \cite{morii} and
\cite{dixon}.

To compare with the DVCS results, let us take the forward limit, i.e.
$-t\rightarrow 0$. Then, from (\ref{LO}), we can see that most of 
the LO helicity amplitudes vanish except
\be
M^{(0)}_{++,++}=2\sqrt{\frac{s+Q^2}{s}},&&
M^{(0)}_{-+,-+}=2\sqrt{\frac{s}{s+Q^2}}.
\en
Since the scalar integrals (\ref{d01}), (\ref{d02}), (\ref{d03}) and
(\ref{c03}) in Appendix are not well-defined at the point $-t=0$, our
methods described above can not be directly applied to the case of
DVCS.  However, this difficulty may be avoided by taking the trace
over the numerator of box integrals. In the DVCS limit ($-t=0$), from
(\ref{kinematic}) we have $p^\prime=\frac{s}{s+Q^2}p$ and the product
of two Dirac spinors can be written as
$u_\alpha(p^\prime,\lambda)\bar{u}_\beta(p,\lambda)\sim(1+\lambda\gamma_5)\rlap/{p}$,
therefore the numerators of the box integrals is of the trace
form. The trace of the numerator can be expressed as a sum of terms
which cancel one of the propagators in the denominator. This can be
done because $k^2$ and $k\cdot p$ can be written as linear
combinations of $k^2$ and $(k+p)^2$, where $k$ is the loop momentum,
and the trace only includes these two type of terms. Since we have
shown that the numerator will be linear combination of two different
denominators, the four-propagator integral will in general become two
three-propagator integrals. After this procedure, we do not need to
carry out the calculation of the four questionable scalar integrals
mentioned above to obtain the DVCS amplitudes, and the calculation
becomes much simpler. After some algebras, we find the NLO results for
the DVCS helicity amplitudes
\be\nn
M^{(1)}_{++,++}&=&M^{(0)}_{++,++}\left(-\frac{\alpha_sC_Ff(\epsilon)}{4\pi}
\right)\left\{\frac{1}{(Q^2)^\epsilon}\frac{1}{\epsilon}
\left[3+2\left({\rm ln}\frac{s}{Q^2}-i\pi\right)\right]\right.\\
&&\left.+9-{\rm ln}\frac{-u}{Q^2} -\frac{2s}{s+Q^2}\left({\rm
ln}\frac{s}{Q^2}-i\pi\right) -\left({\rm
ln}\frac{s}{Q^2}-i\pi\right)^2\right\},\\\nn
M^{(1)}_{-+,-+}&=&M^{(0)}_{-+,-+}\left(-\frac{\alpha_sC_Ff(\epsilon)}{4\pi}
\right)\left\{\frac{1}{(Q^2)^\epsilon}\frac{1}{\epsilon} \left(3+2{\rm
ln}\frac{-u}{Q^2}\right)\right.\\ &&\left.+9-\left({\rm
ln}\frac{s}{Q^2}-i\pi\right) -\frac{2(s+Q^2)}{s}{\rm
ln}\frac{-u}{Q^2}-{\rm ln}^2\frac{-u}{Q^2}\right\}. 
\en
The $1/\epsilon^2$ divergent terms vanish and only $1/\epsilon$
divergence is left, this is due to the fact that there is no soft
divergence in the $-t=0$ limit. Except for an overall normalization
factor, the symmetric and antisymmetric quark amplitudes defined in
\cite{Ji:1998xh} can be written as
\be\label{jxd}
T_q^{(ij)}=M_{i-,i-}+M_{j+,j+},&&T_q^{[ij]}=M_{i-,i-}-M_{j+,j+}.
\en
The CM frame chosen in this paper is different from the CM frame chosen in
\cite{Ji:1998xh}, and the relation between the kinematic variables
used in these two papers can be expressed as
\be\label{sub}
s=\frac{1-x_B}{x_B}\tilde{Q}^2,&&Q^2=2\tilde{Q}^2,
\en
where $\tilde{Q}^2$ represents the variable $Q^2$ defined in
\cite{Ji:1998xh}. After the substitutions of (\ref{sub}), the two
amplitudes derived from (\ref{jxd}) are exactly equal to the
corresponding equations (24) and (25) in \cite{Ji:1998xh} when the
limits of $-t\rightarrow 0$ and $M^2\rightarrow 0$ are taken for the
later case. By checking in the similar way, our results in DVCS limit
are also shown to be in agreement with other existing one-loop results
in \cite{Belitsky:1997rh,Mankiewicz:1997bk,Belitsky:1998wz}.

\section{Conclusion and discussion}

As a complement to the results given in \cite{Diehl:1999tr}, we 
have calculated the NLO QCD corrections to the subprocess helicity 
amplitudes for virtual Compton scattering. The divergent parts, being
consistent with Catani's general formulas, exhibit the same behavior as 
those in electromagnetic form factor and therefore can be attributed
to the form factors introduced in \cite{diehl} which are specific to Compton
scattering. Our finite results can be used to the predictions for
various virtual Compton observables and are compared to the
leading contribution given in \cite{Diehl:1999tr}. One of the
interesting physical quantities is the beam asymmetry for
$ep\rightarrow ep\gamma$
\be
A_L&=&\frac{d\sigma(+)-d\sigma(-)}{d\sigma(+)+d\sigma(-)},
\en
where the labels $+$ and $-$ denote the lepton beam helicity. This
quantity depends on the relative phase between the complex virtual
Compton amplitudes and the real Bethe--Heitler ones. In the LO soft
physics approach, $A_L$ is zero because all amplitudes are real.
However, since the $\alpha_s$ corrections in the photon--parton
subprocess include imaginary parts, $A_L$ may become non-zero in the
NLO soft physics approach \cite{prepare}. This must be a good test for
the handbag mechanism by comparing with the data which will be
obtained in the forthcoming experiments at Jefferson Lab or ELFE-type
accelerators.

\section*{Acknowledgments}
This work started when one of the authors(H.W.H.) was at Kobe
university as the postdoctoral research fellow (No. P99221) of the
Japan Society of the Promotion of Science (JSPS) and H.W.H. would like
to thank the Monbusho's Grand-in-Aid for the JSPS postdoctoral fellow
for financial support. The authors also acknowledge useful comments
from Dieter M\"{u}ller.

\newpage

\appendix
\section{Scalar loop integrals}
The loop integrals from the one-loop graphs of Fig. \ref{fig1} and
Fig. \ref{fig3} can be reduced to a set of 13 scalar integrals which
are given in this Appendix. They all include a common overall
factor
\be
f(\epsilon)=\frac{i\Gamma(1+\epsilon)(4\pi\mu^2)^\epsilon}{16\pi^2}.
\en
Among these 13 integrals, the four-point integrals $D_0^i$ ($i=1-3$)
and the three-point integrals $C_0^i$ ($i=1-6$) are infrared divergent
and ultraviolet finite, while the two-point integrals $B_0^i$
($i=1-4$) include only UV divergence. However, the UV divergent term
containing $\epsilon$ cancels when the two-point functions are
combined to form the tensor integrals and the UV divergences of the
individual graphs also cancel in the sum and, therefore, the final
amplitudes are UV safe.

The four-point functions that we need can be expressed as 
\begin{eqnarray}\nonumber
D_0^1&=&\mu^{2\epsilon}\int\frac{d^nk}{(2\pi)^n}\frac{1}{(k^2+i\varepsilon)
[(p+k)^2+i\varepsilon][(p+q+k)^2+i\varepsilon][(p^\prime+k)^2+i\varepsilon]}
\\\label{d01}
&=&\frac{f(\epsilon)}{st}\left\{\frac{2}{\epsilon^2}
\left(\frac{1}{(-s-i\varepsilon)^\epsilon}+\frac{1}{(-t)^\epsilon}
-\frac{1}{(Q^2)^\epsilon}\right)-\frac{\pi^2}{3}+f_1(s,t,u)\right\},\\\nonumber
D_0^2&=&\mu^{2\epsilon}\int\frac{d^nk}{(\pi)^n}\frac{1}{(k^2+i\varepsilon)
[(p+k)^2+i\varepsilon][(p-q^\prime+k)^2+i\varepsilon]
[(p^\prime+k)^2+i\varepsilon]}\\\label{d02}
&=&\frac{f(\epsilon)}{ut}\left\{\frac{2}{\epsilon^2}
\left(\frac{1}{(-u)^\epsilon}+\frac{1}{(-t)^\epsilon}
-\frac{1}{(Q^2)^\epsilon}\right)-\frac{\pi^2}{3}+f_2(s,t,u)\right\},\\\nonumber
D_0^3&=&\mu^{2\epsilon}\int\frac{d^nk}{(2\pi)^n}\frac{1}{(k^2+i\varepsilon)
[(p+k)^2+i\varepsilon][(p+q+k)^2+i\varepsilon][(q^\prime+k)^2+i\varepsilon]}
\\\label{d03}
&=&\frac{f(\epsilon)}{su}\left\{\frac{2}{\epsilon^2}
\left(\frac{1}{(-s-i\varepsilon)^\epsilon}+\frac{1}{(-u)^\epsilon}
-\frac{1}{(Q^2)^\epsilon}\right)-\frac{\pi^2}{3}+f_3(s,t,u)\right\}.
\end{eqnarray}
where 
\be\nn
f_1(s,t,u)&=&{\rm ln}^2\frac{-t}{s+Q^2}+{\rm ln}\frac{-t}{s+Q^2}{\rm ln}\frac{-u}{s}
-2{\rm ln}\frac{-t}{s+Q^2}{\rm ln}\frac{-t}{s}+{\rm Li_2}\left(\frac{-u}{s+Q^2}\right)\\\nn
&&-{\rm Li_2}\left(1+\frac{Q^2}{t}\right)
-{\rm Li_2}\left(1-\frac{Q^2u}{st}\right)+{\rm Li_2}\left(1+\frac{u}{s}\right)
+{\rm Li_2}\left(\frac{s}{s+Q^2}\right)\\
&&-2i\pi {\rm ln}\frac{-t}{s+Q^2},\\\nn
f_3(s,t,u)&=&{\rm ln}^2\frac{-u}{s+Q^2}+{\rm ln}\frac{-u}{s+Q^2}{\rm ln}\frac{-t}{s}
-2{\rm ln}\frac{-u}{s+Q^2}{\rm ln}\frac{-u}{s}+{\rm Li_2}\left(\frac{-t}{s+Q^2}\right)\\\nn
&&-{\rm Li_2}\left(1+\frac{Q^2}{u}\right)
-{\rm Li_2}\left(1-\frac{Q^2t}{su}\right)+{\rm Li_2}\left(1+\frac{t}{s}\right)
+{\rm Li_2}\left(\frac{s}{s+Q^2}\right)\\
&&-2i\pi {\rm ln}\frac{-u}{s+Q^2}.
\en
For $Q^2>-u$,
\be\nn
f_2(s,t,u)&=&{\rm ln}^2\left(\frac{-t}{u+Q^2}\right)
+{\rm ln}\left(\frac{-t}{u+Q^2}\right){\rm ln}\left(\frac{s}{-u}\right)
-2{\rm ln}\left(\frac{-t}{u+Q^2}\right){\rm ln}\frac{t}{u}\\\nn
&&+{\rm Li_2}\left(\frac{-s}{u+Q^2}\right)
-{\rm Li_2}\left(1+\frac{Q^2}{t}\right)-{\rm Li_2}\left(1-\frac{Q^2s}{tu}\right)
+{\rm Li_2}\left(1+\frac{s}{u}\right)\\
&&+{\rm Li_2}\left(\frac{u}{u+Q^2}\right),
\en
while for $Q^2<-u$,
\be\nn
f_2(s,t,u)&=&{\rm ln}^2\frac{t}{u+Q^2}
+{\rm ln}\frac{t}{u+Q^2}{\rm ln}\frac{s}{-u}
-2{\rm ln}\frac{t}{u+Q^2}{\rm ln}\frac{t}{u}-\frac{\pi^2}{3}
-\frac{1}{2}{\rm ln}^2\frac{-s}{u+Q^2}\\\nn
&&-\frac{1}{2}{\rm ln}^2\frac{u}{u+Q^2}
-{\rm Li_2}\left(\frac{u+Q^2}{-s}\right)
-{\rm Li_2}\left(1-\frac{Q^2s}{tu}\right)
+{\rm Li_2}\left(1+\frac{s}{u}\right)\\
&&-{\rm Li_2}\left(\frac{u+Q^2}{u}\right)
-{\rm Li_2}\left(1+\frac{Q^2}{t}\right).
\en
In the real photon case ($Q^2\rightarrow 0$), we have 
the simplified results
\be
f_1(s,t,u)&=&-{\rm ln}^2\frac{-t}{s}-2i\pi {\rm ln}\frac{-t}{s},\\
f_3(s,t,u)&=&-{\rm ln}^2\frac{-u}{s}-2i\pi {\rm ln}\frac{-u}{s},\\
f_2(s,t,u)&=&-{\rm ln}^2\frac{t}{u}-\pi^2.
\en 

The three-point functions are
\begin{eqnarray}\nonumber
C_0^1&=&\mu^{2\epsilon}\int\frac{d^nk}{(2\pi)^n}\frac{1}{(k^2+i\varepsilon)
[(p+k)^2+i\varepsilon][(p+q+k)^2+i\varepsilon]}\\
&=&\frac{f(\epsilon)}{(s+Q^2)}\left\{
\frac{1}{(-s-i\varepsilon)^\epsilon}-\frac{1}{(Q^2)^\epsilon}\right\}
\frac{1}{\epsilon^2},\\\nonumber
C_0^2&=&\mu^{2\epsilon}\int\frac{d^nk}{(2\pi)^n}\frac{1}{(k^2+i\varepsilon)
[(p^\prime+k)^2+i\varepsilon][(p+q+k)^2+i\varepsilon]}\\
&=&\frac{f(\epsilon)}{s}\left\{\frac{1}{\epsilon^2}
\frac{1}{(-s-i\varepsilon)^\epsilon}-\frac{\pi^2}{6}\right\},\\\nonumber
C_0^3&=&\mu^{2\epsilon}\int\frac{d^nk}{(2\pi)^n}\frac{1}{(k^2+i\varepsilon)
[(p+k)^2+i\varepsilon][(p^\prime+k)^2+i\varepsilon]}\\\label{c03}
&=&\frac{f(\epsilon)}{t}\left\{\frac{1}{\epsilon^2}
\frac{1}{(-t)^\epsilon}-\frac{\pi^2}{6}\right\},\\\nonumber
C_0^4&=&\mu^{2\epsilon}\int\frac{d^nk}{(2\pi)^n}\frac{1}{(k^2+i\varepsilon)
[(k-q)^2+i\varepsilon][(k-q^\prime)^2+i\varepsilon]}\\
&=&\frac{f(\epsilon)}{(t+Q^2)}\left\{
\frac{1}{(-t)^\epsilon}-\frac{1}{(Q^2)^\epsilon}\right\}
\frac{1}{\epsilon^2},\\\nonumber
C_0^5&=&\mu^{2\epsilon}\int\frac{d^nk}{(2\pi)^n}\frac{1}{(k^2+i\varepsilon)
[(p+k)^2+i\varepsilon][(p-q^\prime+k)^2+i\varepsilon]}\\
&=&\frac{f(\epsilon)}{u}\left\{\frac{1}{\epsilon^2}
\frac{1}{(-u)^\epsilon}-\frac{\pi^2}{6}\right\},\\\nn
C_0^6&=&\mu^{2\epsilon}\int\frac{d^nk}{(2\pi)^n}\frac{1}{(k^2+i\varepsilon)
[(p^\prime+k)^2+i\varepsilon][(p^\prime-q+k)^2+i\varepsilon]}
\\&=&\frac{f(\epsilon)}{(u+Q^2)}\left\{
\frac{1}{(-u)^\epsilon}-\frac{1}{(Q^2)^\epsilon}\right\}
\frac{1}{\epsilon^2}.
\end{eqnarray}

The two-point functions are
\begin{eqnarray}\nonumber
B_0^1&=&\mu^{2\epsilon}\int\frac{d^nk}{(2\pi)^n}\frac{1}{(k^2+i\varepsilon)
[(p+q+k)^2+i\varepsilon]}\\
&=&f(\epsilon)\left(\frac{1}{\epsilon}
\frac{1}{(-s-i\varepsilon)^\epsilon}+2\right),\\\nonumber
B_0^2&=&\mu^{2\epsilon}\int\frac{d^nk}{(2\pi)^n}\frac{1}{(k^2+i\varepsilon)
[(p-p^\prime+k)^2+i\varepsilon]}\\
&=&f(\epsilon)\left(\frac{1}{\epsilon}
\frac{1}{(-t)^\epsilon}+2\right),\\\nonumber
B_0^3&=&\mu^{2\epsilon}\int\frac{d^nk}{(2\pi)^n}\frac{1}{(k^2+i\varepsilon)
[(q+k)^2+i\varepsilon]}\\
&=&f(\epsilon)\left(\frac{1}{\epsilon}
\frac{1}{(Q^2)^\epsilon}+2\right),\\\nonumber
B_0^4&=&\mu^{2\epsilon}\int\frac{d^nk}{(2\pi)^n}\frac{1}{(k^2+i\varepsilon)
[(p-q^\prime+k)^2+i\varepsilon]}\\
&=&f(\epsilon)\left(\frac{1}{\epsilon}
\frac{1}{(-u)^\epsilon}+2\right).
\end{eqnarray}

\end{document}